\documentclass[aip,
rsi,
amsmath,amssymb,
reprint,%
]{revtex4-1}

\usepackage{graphicx}%
\usepackage{dcolumn}%
\usepackage{bm}%

\usepackage[utf8]{inputenc}
\usepackage[T1]{fontenc}
\usepackage{mathptmx}
\usepackage{etoolbox}

\usepackage{subcaption}
\usepackage{caption}
\usepackage{todonotes}
\setuptodonotes{inline}
\usepackage{siunitx}
\DeclareSIUnit{\lpmm}{\text{grooves/mm}}
\DeclareSIUnit\bar{\text{bar}}
\usepackage{comment}

\DeclareUnicodeCharacter{0301}{\'{y}}

\makeatletter
\def\@email#1#2{%
 \endgroup
 \patchcmd{\titleblock@produce}
  {\frontmatter@RRAPformat}
  {\frontmatter@RRAPformat{\produce@RRAP{*#1\href{mailto:#2}{#2}}}\frontmatter@RRAPformat}
  {}{}
}%
\makeatother
\begin{document}

\preprint{AIP/123-QED}

\title[]{Ultra-high-vacuum cluster tool for epitaxial synthesis and optical spectroscopy of reactive 2D materials}
\author{M. Dembecki}
\thanks{These authors contributed equally to this work.}%
\affiliation{ 
Walter Schottky Institute and TUM School of Natural Sciences, Technical University of Munich, 85748 Garching, Germany%
}%

\author{J. Schabesberger}%
\thanks{These authors contributed equally to this work.}%
\affiliation{ 
Walter Schottky Institute and TUM School of Natural Sciences, Technical University of Munich, 85748 Garching, Germany%
}%

\author{M. Bissolo}%
\affiliation{ 
Walter Schottky Institute and TUM School of Natural Sciences, Technical University of Munich, 85748 Garching, Germany%
}%
\author{A. Thurn}%
\altaffiliation[Current address: ]{Cavendish Laboratory, University of Cambridge, J. J. Thomson Avenue, Cambridge, CB3 0US, UK%
}
\affiliation{ 
Walter Schottky Institute and TUM School of Natural Sciences, Technical University of Munich, 85748 Garching, Germany%
}
\author{A. Ulhe}%
\affiliation{ 
Walter Schottky Institute and TUM School of Natural Sciences, Technical University of Munich, 85748 Garching, Germany%
}%
\author{P. Avdienko}
\affiliation{ 
Walter Schottky Institute and TUM School of Natural Sciences, Technical University of Munich, 85748 Garching, Germany%
}%
\author{J. Ulrichs}%
\affiliation{ 
Walter Schottky Institute and TUM School of Natural Sciences, Technical University of Munich, 85748 Garching, Germany%
}%
\author{H. Riedl}%
\affiliation{ 
Walter Schottky Institute and TUM School of Natural Sciences, Technical University of Munich, 85748 Garching, Germany%
}%
\author{G. Koblm\"uller}%
\affiliation{ 
Walter Schottky Institute and TUM School of Natural Sciences, Technical University of Munich, 85748 Garching, Germany%
}%
\affiliation{Institute of Physics and Astronomy, Technical University Berlin, 10623 Berlin, Germany}
\author{E. Zallo}
\email{eugenio.zallo@tum.de}
\affiliation{ 
Walter Schottky Institute and TUM School of Natural Sciences, Technical University of Munich, 85748 Garching, Germany%
}%
\author{J.J. Finley}
\affiliation{ 
Walter Schottky Institute and TUM School of Natural Sciences, Technical University of Munich, 85748 Garching, Germany%
}%

\date{\today}%

\begin{abstract}%
The large-area synthesis of high-crystalline-quality two-dimensional (2D) materials is at the core of novel material integration for semiconductor technology. This effort relies on developing fabrication and characterization techniques that can uncover the material's intrinsic properties by preserving its pristine conditions. In this article, we present an all ultra-high-vacuum cluster for the growth using molecular beam epitaxy of 2D semiconductors that are unstable under ambient conditions and optical spectroscopy using low temperature (\qty{20}{\K}) photoluminescence and Raman scattering. The optical chamber of the setup provides micrometer scale spatial resolution and the ability to scan the entire wafer. This enables \textit{in situ} analysis of the structural and optoelectronic properties of as-grown materials in their pristine form, providing rich and reproducible feedback for both fundamental studies and the optimization of scalable 2D material growth toward integration in advanced devices.
\end{abstract}

\maketitle

\section{Introduction}
Two-dimensional (2D) materials beyond graphene, including black phosphorus, transition metal dichalcogenides, and post-transition metal monochalcogenides (PTMCs), possess unique optical and electronic properties and hold large promise for optoelectronic applications due to their non-zero electronic bandgap~\cite{cui_versatile_2019,long_progress_2019}. However, most of them are prone to reacting with air, making it difficult to assess pristine material properties that are not preserved during oxidation~\cite{wang_chemical_2019}. Ambient pressure passivation of the surface via van der Waals (vdW) encapsulation has been demonstrated to prevent surface degradation. However, such approaches modify the electronic properties and affect the reliability of results by introducing interface states and altering the dielectric environment~\cite{han_effects_2019}. 
To achieve high-quality, uniform 2D films that surpass the limits of mechanical exfoliation of bulk crystals~\cite{novoselov_electric_2004,mak_atomically_2010}, bottom-up epitaxial techniques such as chemical vapor deposition (CVD) and molecular beam epitaxy (MBE) are widely employed~\cite{liu_understanding_2024, zallo_two-dimensional_2023, pianetti2025, rhodes_disorder_2019, Bissolo2025}. Raman and photoluminescence (PL) spectroscopy stand out as non-invasive tools for semiconductor analysis. While the first yields information on crystallinity, phase, and thickness~\cite{cong_application_2020}, the latter probes a wide range of electronic properties from band gap to absorption coefficient, unveils the presence of excitons, and provides valuable insights into optically active defects and vdW heterostructures~\cite{shree_guide_2021}. 
Recently, advanced ultra-high-vacuum (UHV) compatible setups for the study of 2D materials were developed by connecting fabrication and analytical methodologies in a single UHV environment. For example, epitaxial synthesis has been combined with scanning probe microscopy and electron spectroscopy~\cite{bradford_epitaxy_2024}, sputtering with mechanical exfoliation and scanning probe microscopy~\cite{wang_clean_2023}, or Raman spectroscopy with electrical transport~\cite{Shchukin2024}. However, the combination of all-UHV epitaxial growth and optical spectroscopy to understand the pristine properties of high-quality 2D materials is still lacking. 
Herein, we report the assembly of an all-UHV cluster tool for the MBE synthesis of vdW-bonded materials, connected via a UHV transfer channel to an analytic chamber with capabilities for \textit{in situ} optical confocal spectroscopy at sample temperatures between \qtylist{20; 300}{\kelvin}. The system provides access to the ultrapure electronic, optical, and interface properties of the as-grown materials via micro-PL and micro-Raman mapping of large-scale (up to \qty{3}{\cm} diameter) samples with a resolution of \qty{1.1}{\um} (\qty{18.9}{\um}) at 300 K (20 K). We demonstrate the growth and in situ optical spectroscopy of preserved PTMCs, highlighting their stability over an extended period, both with and without illumination. 
This integrated approach provides reliable access to as-grown vdW-bonded materials under pristine UHV conditions, enabling systematic and contamination-free investigations of the intrinsic properties of novel 2D materials.

\section{Experimental Setup: The synthesis-analytic cluster Tool \label{setup}}

\begin{figure*}
    \includegraphics[width=0.95\linewidth]{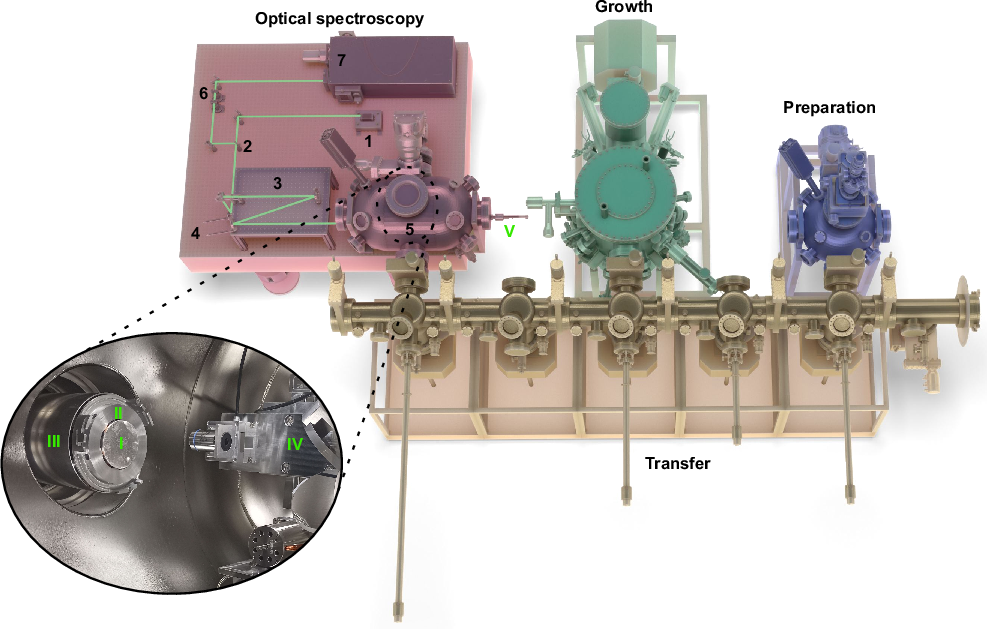}
\caption{CAD rendered model of the full cluster. The cluster consists of the transfer system (light beige), the preparation chamber (blue), the molecular beam epitaxy (MBE) growth chamber (green), and the optical spectroscopy chamber (red). The components of the optical setup are indicated with numbers: laser (1), beam splitter (2), second level aligned with the optical spectroscopy chamber (3), beam steering mirror (4), objective inside the analytic chamber (5), spectral filters (6), and spectrometer with charge-coupled device  (7). The inset shows a photograph depicting the interior of the analytic chamber with the main components consisting of an indium foil (I) covering the copper piece, a 3" molybdenum ring designed to receive the sample holder (II), a cryo shroud made from aluminum (III), and the x-y-z objective stage (IV). A linear feedthrough is used to move the microscope (V, see the main figure). \label{fig:cluster}}
\end{figure*}

As shown in Fig.~\ref {fig:cluster}, the cluster comprises three main chambers for sample preparation, growth, and analysis, shaded in blue, green, and red, respectively. These chambers are connected with a modular Riber modutrac system for sample transfer in UHV, highlighted in beige. A sample typically passes through the system in the given order before being taken out of UHV for ex-situ analysis and processing. All chambers were adapted to fit the Riber 3" substrate holder standard. 

\begin{figure}[h]
    \includegraphics[width=0.45\textwidth]{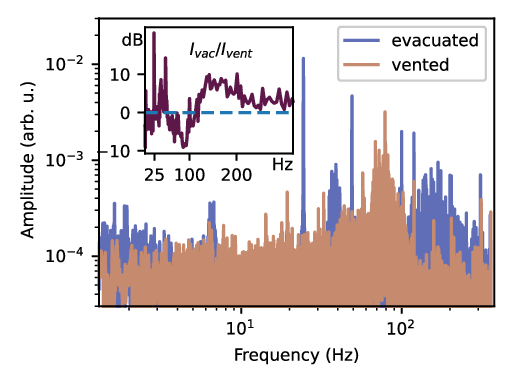}
    \caption{Vibration frequency spectrum before and after evacuating the bellow connecting the chamber to the cluster. The spectra are obtained from the Fourier transform of the laser spot position reflected from a polished SiO$_2$ surface, recorded using high-frequency video imaging. The inset shows the ratio of the vibration amplitudes of both spectra in dB.\label{fig:Vibration}}
\end{figure}

To remove contaminants from the substrate prior to growth, the preparation chamber is equipped with a heater capable of reaching \qty{1100}{\kelvin}, as well as a hydrogen cracker cell for chemical cleaning. To monitor the sample outgassing, a residual gas analyzer (Stanford Research Systems, RGA 200) is installed. A base pressure of \qty{2e-10}{\milli\bar} is achieved with a turbomolecular pump. 
The growth of 2D semiconductors takes place in the adjacent MBE chamber (Riber Compact 21 EB200M), which is equipped with an electron gun and a fluorescent screen for growth monitoring via reflection high-energy electron diffraction (RHEED), a quadrupole mass spectrometer, and a liquid nitrogen-cooled cryo shroud to decrease the background pressure by preventing re-evaporation of molecules from the chamber walls. The chamber is equipped with standard Knudsen effusion cells for gallium, indium, and aluminum, two valved cracker cells for selenium and tellurium (Riber S.A., VCOR 110), and a nitrogen plasma cell (VEECO Inc., RFS-N). Multi-pocket electron beam evaporators (MBE-Komponenten GmbH, EBVM 200) expand the range of PTMC alloys by enabling the evaporation of low-vapor-pressure materials. Before the start of material synthesis, a base pressure of \qty{5e-11}{\milli\bar} was reached using an ion getter pump. \\
The grown sample can be transported to the analytic chamber for visual inspection and optical spectroscopy. The inset in Fig.~\ref{fig:cluster} shows a photograph of the interior of the optical spectroscopy chamber. The sample holder (Riber bayonet system, typically a 3" molybdenum ring on which a 2" wafer is held in place by a sheet metal ring) is cooled down by bringing it in contact with the cold finger of a pulse tube cooler (Sumitomo (SHI) Cryogenics of America, CH-204 6.5K) by means of an oxygen-free high conductivity copper based adapter. The adapter surface is covered with a thin layer of indium foil (I) to improve thermal contact and prevent copper contamination in the growth chamber. A molybdenum ring (II) with two sets of recipients for the Riber bayonet transfer system encloses the copper piece. An aluminum-based radiation shield (III) is mounted to the first stage of the cryostat to reduce thermal load to the second stage. The optical assembly inside the chamber (IV) consists of a 40$\times$ microscope objective mounted on a x-y-z stage powered by UHV-compatible stepper motors, along with a piezo nano-positioning system in the x-y direction. The laser beam for the free-space excitation (see the green path depicted in Fig.~\ref{fig:cluster}) enters the analytic chamber from the left through a quartz viewport of deep-UV quality and is deflected to the sample via a custom-made periscope with two 3" mirrors placed in UHV on a cart. To facilitate sample transfer between the modutrac and the cold finger, the microscope tower and periscope are mounted on a cart that can be moved between fixed positions by a push-pull transfer arm on the right (V). The vacuum environment can be monitored with a second residual gas analyzer. A base pressure of \qty{8e-9}{\milli\bar} is reached with a turbomolecular pump. Note that this value is limited by the large, complex surface area and by the use of polymers for electrical insulation, which also limit the maximum baking temperature to \qty{80}{\degreeCelsius}. \\
To isolate the chamber from external vibrations, the chamber is placed on an optical table that supports the optical setup. The connection to the cluster is made via a flexible bellow, which allows the active dampening of the table to function effectively (the bellow is supported by two metal rods to prevent compression). This approach suppresses the low-frequency components, even though some vibrations are transmitted from the cluster to the microscope. 
This is illustrated in Fig.~\ref{fig:Vibration}, where the Fourier transform of the laser beam reflection position on the SiO$_2$ surface, obtained from video imaging at \qty{719}{\hertz}. In comparison between the vented and evacuated state of the connecting bellow, one observes that stronger coupling to the cluster in the evacuated case leads to vibration increases exceeding \qty{10}{\deci \bel} at distinct frequencies of \qty{25}{\hertz} and \qty{50}{\hertz}, as well as weaker features at higher harmonics of \qty{25}{\hertz}). Additionally, broader bands centered around \qty{40}{\hertz} and \qty{150}{\hertz} exhibit increases of approximately \qty{7}{\deci \bel}. The sharp features were identified as originating from membrane pumps on the cluster and can be mitigated by damping their vibration transmission. In contrast, the broader features are attributed to resonances of the optical assembly and vacuum chamber system. Strong coupling to the cluster may shift this resonance from about \qty{80}{\hertz} in the vented state to above \qty{100}{\hertz} after evacuation. Below \qty{10}{\hertz}, only a minor increase is observed, approaching the sensitivity limit of the measurement technique at around \qty{2}{\hertz}. As the timescale of spectral measurements is >\qty{10}{\second}, the effective spatial resolution is not significantly altered as characterized later in section \ref{sec:fullscan}.\\
The optical setup outside of the UHV chamber consists of two levels. On the lower level, the beam from a narrow-band 532~nm laser (Hübner Photonics GmbH, Cobolt-08 DPL, see Fig.~\ref{fig:cluster}) is first filtered spectrally (OptiGrate Corp., BNF) and spatially to maintain a collimated Gaussian profile. The excitation and detection paths are separated by a wedged ND filter (2). The light coming from the sample (3) is again spatially filtered before the laser line is suppressed using volumetric Bragg gratings (6). The signal then passes into a spectrometer (Quantum Design GmbH, Shamrock 750) of \qty{750}{\mm} focal length (7) with three gratings of \num{300}, \num{600}, and \qty{1200}{\lpmm}. The diffracted light is collected by an open-electrode charge-coupled device (CCD) (Quantum Design GmbH, iDus-420-OE), cooled thermoelectrically to -80~K. Together with a spectrometer, a CCD, and with an optimized entrance slit, the three gratings yield a resolution of \num{0.21}, \num{0.1}, and \qty{0.05}{\nm}, respectively.
Aligned to the height of the lower periscope mirror inside the analytic chamber, an upper level of the optical setup (3) contains the steering mirror used to scan the excitation laser beam (4) as well as the imaging system as parts of the combined optical path.

\section{Optical Spectroscopy Chamber: Performance Tests}
To provide high-resolution spectroscopic information on the as-grown 2D semiconductor at the wafer scale, several demands must be met.
Wafer-scale Raman spectroscopy can map how variations in growth conditions influence structural properties, such as stoichiometry, crystal phase and growth mode, and help identify the boundaries between different phases, as demonstrated in the study of the epitaxial growth of PTMC gallium selenide (GaSe) on sapphire~\cite{Bissolo2025GaSe}. Furthermore, these measurements can help optimizing the homogeneity of the epitaxial layer, which remains a significant challenge in the scalable fabrication of 2D materials~\cite{zallo_two-dimensional_2023, Bissolo2025}. To probe local variance in morphology or substrate-induced strain fields~\cite{Liu2020_GaSeStrainMBE}, micrometer resolution scanning is necessary.
Local effects, like defects~\cite{bissolo2025dlts}, grain boundaries, and hetero- and homo-interfaces, also have a strong effect on PL emission~\cite{Chow2015_TMDDefectPL, ly2014_PLGrainBoundaries, rakhlin2024_GaSe/Ga2Se3MBEPL} but typically only emerge at low temperatures. PTMCs are reported to have strong electron-phonon coupling, especially in the few-layer limit~\cite{Le2024_RamanPL, Paylaga2024_InSePl, Lugovskoi_InSeTheory}, and exciton binding energies of around \qty{20}{\meV} in bulk material \cite{zalamai_2020, Shubina_2019}. To unravel these fundamental properties, control of sample temperature down to a few K is needed. 
Finally, the presence of oxygen can lead to degradation of GaSe over time, a process that is accelerated under the illumination of light~\cite{Murray2025_OxidationStatistically, Smiri2024_OxidationLongTerm, Bergeron17GaSeOxidation}.\\ 
The following sections discuss full wafer scanning at room temperature with micrometer resolution, the influence of vibrations generated by the closed-cycle cryostat on the spatial resolution of the setup, cooling of the sample to approximately \qty{20}{\kelvin}, and the preservation of samples in the UHV environment of our cluster to rule out any (photo-)oxidation. 

\subsection{\label{sec:fullscan}Full wafer scanning with micrometer resolution}

\begin{figure}[h]
    \includegraphics[width=0.45\textwidth]{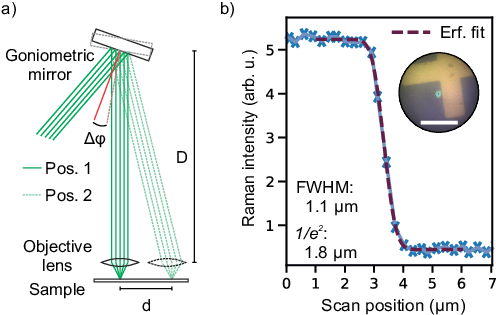}
    \caption{a) Geometric optics sketch of the employed pseudo-4f laser scanning technique at two exemplary and exaggerated angles. Tilting the mirror by $\Delta\varphi$ shifts the spot on the sample by a distance d. Since the distance between the goniometric mirror and the objective lens (D) is \qty{1.3}{\m}, a small angle of $\Delta\varphi$ = 1.1$^\circ$ is sufficient to access the full sample. b) Spot size determination using the intensity of the \qty{520}{\per\centi\meter} mode of Si during a line scan across an Au marker. The dashed line indicates a fitting model based on the error function to determine the width of the transient as a measure of the Gaussian beam width. The inset depicts the cross-shaped marker and laser spot in the in-situ imaging. (Scale bar is \qty{10}{\um}) \label{fig:Pseudo4f}}
\end{figure}

To enable wafer-scale scanning under the constraints of the sample holder transfer system in UHV and the required cryostat configuration, we developed a solution where the objective is scanned over the sample surface and tracked by the excitation/detection beam path. The UHV-compatible microscope tower (see section~\ref{setup}) has a travel distance of \qty{52}{\mm} in the x- and y-directions, combined with sub-nm precision in position accuracy in a \qtyproduct{100 x 100}{\um} area using piezoelectric actuators. As depicted in Fig.~\ref{fig:Pseudo4f}a, the beam steering is obtained by employing a pseudo-4f scanning technique by means of a single planar mirror outside the chamber, approximately D = \qty{1.3}{\m} away from the objective. This approach requires only a goniometric mirror and the objective for confocal imaging across the full sample surface, taking advantage of the long distance between the mirror and the objective. In our case, tilting the mirror by $\Delta\varphi$~=~0.55$^\circ$ results in a scanned distance of d~=~\qty{25}{\mm} on the sample. The objective tracks the beamline, including a small compensation to account for the slight change in the angle of incidence, thereby maintaining a collinear beam path for both the incoming and outgoing beams and preserving the condition for confocal microscopy. This follows the same principle as conventional 4f scanning, with the key difference that the two lenses used to map to/from the Fourier plane are omitted. Clearly, scanning within the field of view also occurs in the pseudo-4f case, leading to minor positioning inaccuracies of about 0.1\% relative to the distance traversed. These inaccuracies can be readily corrected through calibration.

Fundamentally, the spatial resolution is limited by the spot size, which was determined for the \qty{532}{\nm} laser by scanning the focal spot over the edge of a gold marker on a silicon substrate and measuring the intensity of the reflected light using a power meter as well as the intensity of the \qty{520}{\per\centi\meter} mode of Si in the spectrometer, as depicted in Fig.~\ref{fig:Pseudo4f}b. By fitting the transient with an error function to model the underlying Gaussian beam shape, we obtain a full width half maximum (FWHM) of \qty{1.2}{\um} and \qty{1.1}{\um} at room temperature for the two detection methods, respectively, independent of the scanning direction. 

\begin{figure}[h]
    \includegraphics[width=0.45\textwidth]{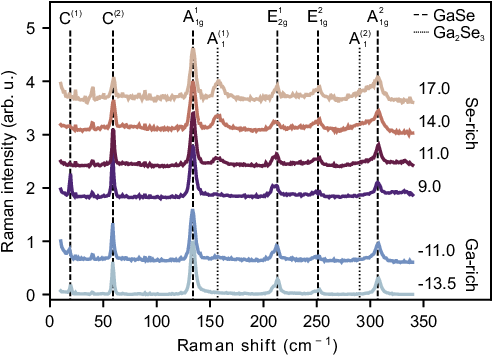}
    \caption{Waterfall plot of Raman spectra across an as-grown 2" GaSe layer on sapphire using a \qty{532}{\nm} laser excitation with a power of \qty{280}{\uW}. The numbers at the right of the spectra denote the respective measurement positions in the y-direction in millimeters from the wafer center. This corresponds to moving from gallium-rich growth conditions (negative values) to selenium-rich conditions (positive values) along a center axis. The dashed and dotted lines indicate the expected peak positions of Raman modes in GaSe with 1:1 and 2:3 stoichiometry, respectively. C$^{(\text{1})}$ and C$^{(\text{2})}$ are shear modes. All Raman modes were identified using literature values, see refs~\cite{rakhlin2024_GaSe/Ga2Se3MBEPL, Claro_GaSeMBE2023, Shiffa_MBEGaSe2023, Osiekowicz_Raman2021, yamada1992_ga2se3}. \label{fig:Raman_linescan}}
\end{figure}

A typical result of a wafer-scale Raman spectroscopy measurement of grown GaSe on sapphire substrate is presented in Fig.~\ref{fig:Raman_linescan}. The normalized Raman spectra were recorded from GaSe grown in the MBE chamber on a 2" sapphire substrate in the wavenumber range between \qty{15} and \qty{340}{\per\cm}. The spectra were taken at different positions on two opposite sides of the wafer where the Ga/Se flux ratio was very different during the sample growth. This flux variation across the wafer arises since the growth was performed under non-rotating conditions to intentionally induce a large gradient in Ga/Se flux ratio across the entire wafer~\cite{Bissolo2025GaSe}. Modes associated with a Ga/Se flux ratio of 1:1 are indicated with dashed lines~\cite{Hoff1975_Raman, Le2024_RamanPL} and are present at all positions. The additional peak appearing at \qty{155}{\per\cm} for positions y $\geq$ \qty{11}{\mm} from the center of the sample as well as the shoulder at \qty{292}{\per\cm} correspond to the $\textrm{A}_\textrm{1}$ mode of $\textrm{Ga}_\textrm{2}\textrm{Se}_\textrm{3}$~\cite{yamada1992_ga2se3}, indicating the coexistence of both phases in the stoichiometric ratios of 1:1 and 2:3 in this region of the wafer. The integrated intensity of the \qty{155}{\per\cm} mode relative to the $\textrm{A}^\textrm{1}_\textrm{1g}$ mode at \qty{134}{\per\cm} decreases from 0.93 at y~=~\qty{17}{\mm} to 0.37 at y~=~\qty{11}{\mm}, following the gradient in flux ratio. For positions of y~=~9~mm and y~=~-11~mm, the $\textrm{A}_\textrm{1}$ peak of $\mathrm{Ga_2Se_3}$ is much weaker at ratios below 0.1, indicating less growth of the 2:3 stoichiometry. Under more Ga-rich conditions at y~=~-13.5~mm, the material exhibits no signature of the $\mathrm{Ga_2Se_3}$ phase~\cite{Bissolo2025GaSe}. \\
The use of volumetric Bragg grating notch filters enables the observation of low-wavenumber Raman modes down to $\sim$\qty{15}{\per\cm}, as visible in Fig.~\ref{fig:Raman_linescan}. In GaSe, the \qty{20}{\per\cm} mode is known to shift towards the Rayleigh line and split into different modes depending on the exact layer number for thicknesses below $\approx10$ layers~\cite{Longuinhos2016_RamanGaSe, Molas2021, Lim2020_GaSeRaman}. This behavior can be analyzed to determine the layer thicknesses of grown material in the few-layer regime and is within the capabilities of our cluster tool, but is not presented here.
Therefore, the micro-Raman spectroscopy presented confirms that the UHV optical spectroscopy chamber is capable of detecting  different material compositions and phases, due to variations in growth parameters on the scale of a wafer, as well as local variations from non-coalesced islands on the micrometer scale. All these measurements can be done in the UHV environment, without the wafer having been exposed to ambient conditions.

\subsection{\label{sec:PL} Temperature control}

\begin{figure}
    \includegraphics[width=0.45\textwidth]{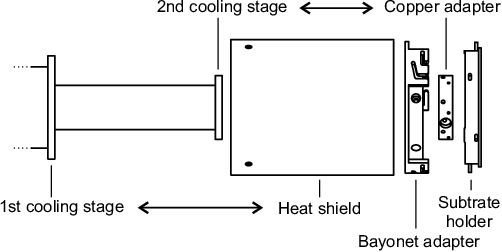}
\caption{Enlarged schematic of the cryostat and its attached components. The heat shield is mounted to the first cooling stage, while the copper adapter is attached to the second stage. The bayonet adapter is radially fixed to the copper adapter. As shown, the bayonet mechanism is located at the top and bottom of the adapter and is designed to accommodate the substrate holder in two positions. The position closer to the cold finger provides thermal contact for the substrate. \label{fig:cryo_drawing}}
\end{figure}

\begin{figure}[b]
    \includegraphics[width=0.45\textwidth]{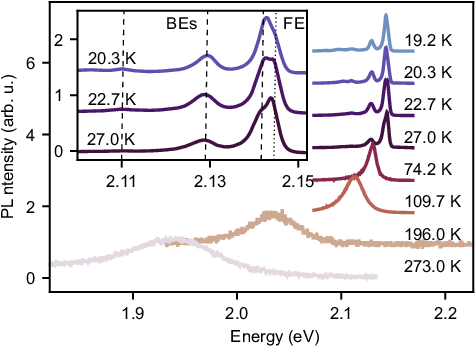}
    \caption{Temperature dependent normalized photoluminescence spectra of epitaxially grown $\gamma$-$\mathrm{In_2Se_3}$ on sapphire in the temperature range 20--300 K. Data was recorded using a \qty{532}{nm} excitation laser with a power of \qty{3}{\micro\watt} and an integration time of \qty{20}{s}. The inset highlights the evolution of bound (BE) and free excitons (FE) as a function of the temperature control near the minimum temperature. The dashed and dotted lines indicate the position of the emission lines of bound excitons and the free exciton, respectively. \label{fig:temperature_series} }
\end{figure}

Ensuring reliable thermal coupling between the sample and the cold finger is typically achieved by either using an adhesive with high thermal conductivity or by high contact pressure using bolts or springs~\cite{Dhuley2019_ThermalContact, SALMON2022_ThermalContact}. In our case, neither of the two options is possible due to the geometry dictated by the sample holder (see Section~\ref{setup}) and the all-UHV environment. Any additional mechanism to exert more external force would compromise sample integrity or the transfer process. 
In order to efficiently cool down the sample under these constraints, several measures were developed, as shown in the inset of Fig.~\ref{fig:cluster} and in schematic form in Fig.~\ref{fig:cryo_drawing}. First, the copper adapter connecting the cold finger to the growth substrate was minimized in length (\qty{10}{\mm}), while accommodating the heater (\qty{6.35}{\mm}) and the temperature sensor (\qty{8.2}{\mm}). The socket receiving the temperature sensor is located below the center of the copper adapter in Fig.~\ref{fig:cryo_drawing}. Then, a heat shield made of aluminum was attached to the first stage of the cryostat, decreasing the radiative heat load to the colder second stage. Finally, a thin indium foil was placed between the cold finger and the copper adapter, as well as on top of the adapter, to substantially increase the contact area at both interfaces~\cite {SALMON2022_ThermalContact}.
Accurate determination of the sample temperature is crucial for reliable analysis of optical spectroscopy data. Therefore, we carefully calibrated the temperature of different wafer materials - silicon, GaAs, and sapphire - to the temperature of the cold finger as measured by a silicon diode, which serves as a reference during regular operation. The temperature of each sample was determined electrically using a resistance sensor attached to the substrate (not shown) and, for an optically active substrate, by extracting the temperature dependence of the band gap from PL spectra. Without heating, the lowest temperatures reached for a GaAs wafer are $T_\text{sample}$ = \qty{20.3}{\kelvin} and $T_\text{cold finger}$ = \qty{7.8}{\kelvin}, respectively. The temperature offset $\Delta$$T$~=~$T_\text{sample}$ $-$ $T_\text{cold finger}$ decreases to \qty{2}{\kelvin} for temperatures above \qty{80}{\kelvin}. The base sample temperature varied between \qty{19.2} and \qty{27.3}{\kelvin} for sapphire and silicon, respectively. This variation arises from differences in the surface absorption and emissivity of black body radiation~\cite{Franta2018_Si, querry1985optical} and the thermal conductivity~\cite{glassbrenner1964thermal, Touloukian_1971_ThermalConductivityNonMetals} of the substrate materials. The presence of a native oxide on the Si wafer with a comparably higher absorption coefficient~\cite{Franta2017_SiO2} and lower thermal conductivity~\cite{Touloukian_1971_ThermalConductivityNonMetals} is likely to increase the base temperature achievable. In addition, the surface roughness at the interface with the cold finger varies depending on the wafer's manufacturing method.
Based on the previously established full-scale calibration, we performed a temperature-dependent PL series on MBE-grown $\gamma$-$\mathrm{In_2Se_3}$. The investigated sample consists of a \qty{200}{\nano\meter}-thick $\gamma$-$\mathrm{In_2Se_3}$ layer grown on sapphire using an In/Se flux ratio of \qty{30} and a substrate temperature of \qty{530}{\degreeCelsius}. The evolution of PL intensity was monitored over temperatures between \qty{19.2} and \qty{273}{\kelvin}. Typical results are presented in Fig.~\ref{fig:temperature_series}. The magnified spectra of the lowest temperatures (see inset) exhibit four peaks at \qty{20.3}{\kelvin}, which are centered at \qty{2.110}, \qty{2.129}, \qty{2.144}, and \qty{2.145}{\eV}, respectively. 
A reduction in the strength of the PL signal of all peaks relative to the highest-energy one (centers at \qty{2.144}, and \qty{2.145}{\eV}) can be observed with increasing temperature. This supports the attribution of the highest-energy feature to a free exciton (FE) and the others as bound excitons (BEs), as reported in the literature~\cite{Ohtsuka2000_In2Se3PL, lyu2010, Brucker2022_In2Se3PLAbsorption}. The thermal dissolution of BEs in $\gamma$-$\mathrm{In_2Se_3}$, manifesting in an exchange of the dominant peak in the PL spectrum, takes place within a temperature range of $<$ \qty{10}{\kelvin} (see inset), which can precisely be set and maintained in our setup.

\subsection{Scanning capabilities at cryogenic temperatures}

\begin{figure*}[ht]
    \includegraphics[width=0.82\textwidth]{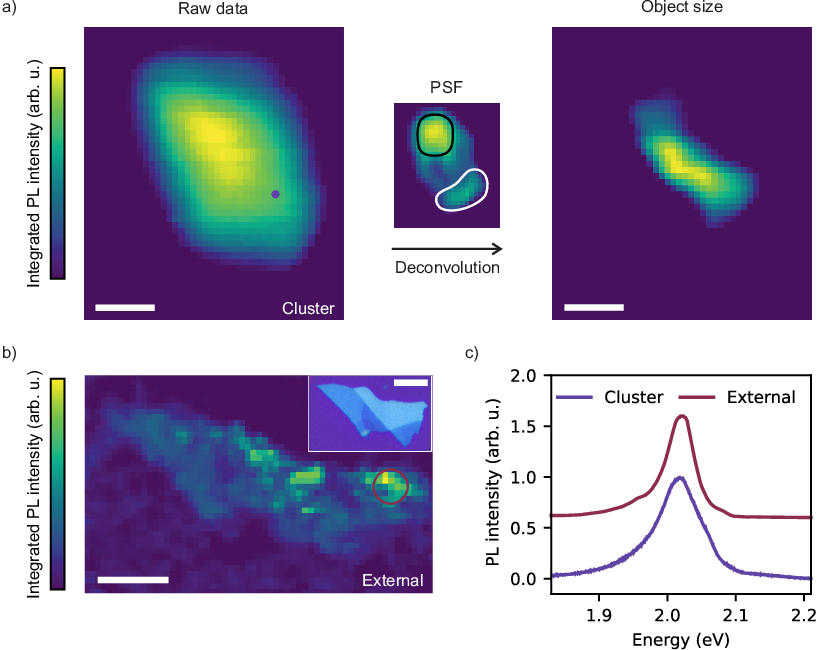}
    \caption{a) Application of the deconvolution approach to the emission from a mechanically exfoliated GaSe sample. The raw PL map (left) is deconvolved using the point-spread function (PSF, center) to produce a more resolved final map (right). The two main spots in the PSF are indicated in black and white. Scale bar: \qty{10}{\um}. b) The PL map of the same sample measured in a flow cryostat without vibration. The inset shows an optical micrograph of the sample. Scale bars: \qty{10}{\um}. c) Normalized spectra from spots in similar positions in both measurement setups. The spots are shown in the respective colors in (a) and (b). All measurements were performed with a \qty{532}{\nm} laser and excitation powers of \qty{200}{\uW} (\qty{100}{\uW}) for the PL map in the cluster (external setup), and \qty{0.2}{\uW} for the quantum dot point source. \label{fig:deconvolution}}
\end{figure*}

The spatial resolution of the optical setup operated under cryogenic conditions was measured using a GaSe sample fabricated via the dry-transfer method~\cite{CastellanosGomez2014_DryTransfer} and encapsulated in hBN (both GaSe and hBN are commercial materials from 2Dsemiconductors Inc.), which protects the material from degradation in air while allowing for external study of morphology and spectral properties. Figure~\ref{fig:deconvolution}a (left) shows the PL map (raw data) of the sample recorded in the optical spectroscopy chamber of the cluster, but no clear contours or details are visible. The image has a size of \qtyproduct{43 x 28}{\um}. However, when the PL map is measured in a helium flow cryostat (spatial resolution $\approx$\qty{1}{\um} and low vibrations) from the same device with the sample oriented in the same direction, the outline of the flake, as well as individual bright spots, are clear (see Figure~\ref{fig:deconvolution}b). The lower resolution of the PL data measured in the cluster is caused by the horizontally mounted compressor, which pumps helium through the cold head at a frequency of \qty{2}{\Hz}. This results in non-harmonic motion of the sample in all three spatial dimensions (i.e., in the focal plane and perpendicular to it), which can be understood as an increase in the effective size of the laser spot.
The spatial resolution of this all-UHV PL setup was quantified through its point spread function (PSF), which included the vibrational contribution. For this purpose, we measured the PL emission from a point-like emitter over a three-dimensional grid of coordinates. We selected a spatially and spectrally isolated InAs/GaAs quantum dot (QD) emitting at \qty{937}{\nm} from a low-density region of a sample grown by MBE via a self-assembly process~\cite{Trotta2012}. The resulting spatial distribution for a single focus position is shown in the center of Fig.~\ref{fig:deconvolution}a (see PSF). The overall shape is approximately elliptical, consisting of a circular spot at the top left (black contour) and a second elongated spot at the bottom right (white contour). The centers of the two regions are \qty{12.5}{\um} apart, with the FWHM of the black contour of \qty{8.9}{\um}. These two regions are each maximal in intensity for different focus positions \qty{8}{\um} apart, a distance significantly larger than the depth of field ($\approx$ \qty{2}{\um}). Scanning of the edge of a gold marker and measuring the reflected power, as described in section \ref{sec:fullscan}, yields a FWHM of \qty{8.9}{\um} and \qty{18.9}{\um} along the short and long half-axis of the ellipse, respectively, which is in agreement with the size determined from the direct imaging. \\
By using the PSF as a kernel, we can apply the Richardson-Lucy deconvolution~\cite{Richardson:72, lucy74, dey2004} to the 'raw data' map, thus revealing more information about its size and shape. Figure~\ref{fig:deconvolution}a (right, 'object size') shows the PL map after the deconvolution, which has a size of \qty{26}{\um} $\times$ \qty{11}{\um}. The deconvoluted map matches well in shape and size with the high-resolution map in Fig.~\ref{fig:deconvolution}b as well as the inset showing an optical image. It is essential to note that only statements about the sample's size and shape can be inferred, as spectral information at each spatial position is lost during the mathematical process. However, the emission energy of specific regions can still be extracted by integrating only a spectrally narrow region.
Figure~\ref{fig:deconvolution}c compares the normalized PL spectra of two spots (see the blue and red markers on the maps of Figs.~\ref{fig:deconvolution}a,b, respectively) measured in different setups. The peak energies are \qty{2.024}{\eV} in the cluster and \qty{2.023}{\eV} in the external setup with a FWHM of \qty{75}{\meV} and \qty{49}{\meV}, respectively. The difference in FWHM is likely due to the larger spot size. Overall, the spectra match well in terms of peak shape and energy, considering the vibrations and their influence. 
Finally, we have shown that cryostat vibrations limit the resolution of PL measurements; however, image-processing methods can mitigate this loss to some extent. In a proof-of-principle experiment, the PL emission energy and line shape in external measurements can also be matched with spectra taken in our setup. It should be noted that a sufficiently homogeneous sample with uniform regions of several \unit{\um^2} is necessary for this approach in order to ensure that a given emission energy is not simply the result of averaging over many different emissions.

\subsection{Conservation of reactive 2D materials}

\begin{figure*}[ht]
    \includegraphics[width=0.95\textwidth]{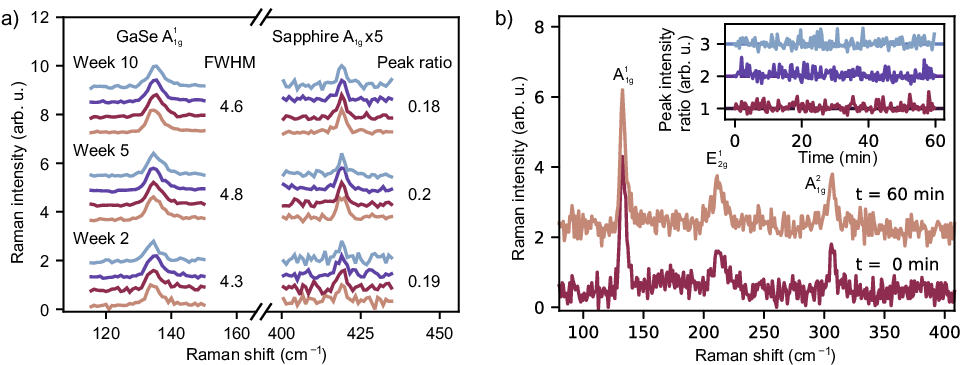}
    \caption{Stability of the grown GaSe material in the cluster. a) $\textrm{A}_\textrm{1}$ mode of Raman spectrum from GaSe on sapphire after 2, 5, and 10 weeks in vacuum, measured at the same four spots within an area of \qtyproduct{10 x 100}{\um} area. The average FWHM of the $\textrm{A}^\textrm{'}_\textrm{1}$ mode, as well as the average ratio of the integrated intensities of the two peaks, are indicated for each timestep. b) Raman spectrum of a randomly chosen point before and after \qty{1}{\hour} of illumination. The inset depicts the evolution of the peak intensity ratio relative to the initial value $(\textrm{I}(\textrm{A}_\textrm{1g})_{\textnormal{sapphire}}/\textrm{I}(\textrm{A}^\textrm{1}_\textrm{1g})_{\textnormal{GaSe}}) / (\textrm{I}(\textrm{A}_\textrm{1g})_{\textnormal{sapphire}}/\textrm{I}(\textrm{A}^\textrm{1}_\textrm{1g})_{\textnormal{GaSe}})(t=0)$ for three spots over the course of \qty{1}{\hour}. The measurements were taken with a \qty{532}{\nm} laser excitation wavelength and a power of \qty{140}{\uW} and  \qty{215}{\uW} for a) and b), respectively. \label{fig:Raman_Stability}}
\end{figure*}

Under ambient conditions (i.e., in the presence of oxygen), certain PTMCs such as GaSe react by forming crystalline $\mathrm{Ga_2Se_3}$, amorphous $\mathrm{Ga_2O_3}$, and amorphous Se. The compound products form a layered film on the GaSe surface, limiting further oxidation, whereas Se forms clusters \cite{liu2022, hong2022}. The presence of water vapor leads to a similar degradation of this type of semiconductor \cite{Bergeron17GaSeOxidation, Kowalski2019_HumidityOxidation}. The oxidation was demonstrated in several reports (see Refs.~\cite{zhao2018_oxidation, delpozozamudio2015, Murray2025_OxidationStatistically, Smiri2024_OxidationLongTerm}). For example, in Raman measurements with the emergence of peaks characteristic of all three products as well as a decrease in the intensity of the $\textrm{A}^\textrm{1}_\textrm{1g}$ mode of GaSe. The typical timescales of exposure to ambient conditions range from eight hours~\cite{Murray2025_OxidationStatistically}, over \qtyrange{100}{200}{\hour}~\cite{zhao2018_oxidation, delpozozamudio2015}, to several weeks~\cite{Smiri2024_OxidationLongTerm}. Illumination accelerates the degradation process~\cite{Bergeron17GaSeOxidation} with a prominent Raman signature of $\alpha$-Se reported as quickly as after \qty{60}{\second} when illuminated with a power density of \qty{0.11}{\mW/\um^2} at \qty{532}{\nm}~\cite{zhao2018_oxidation}.
The UHV pressure range of our cluster (\qtyrange{1e-10}{1e-8}{\milli\bar}) with an even lower partial pressure of oxygen (\qty{5E-10}{\milli \bar}) allows us to preserve the sample from degradation and, thereby, access the pristine material properties. Based on this, we have conducted an experiment to demonstrate the possibility of virtually indefinite storage of PTMC samples in our cluster without affecting their material properties. Fig. \ref{fig:Raman_Stability}a shows the characteristic $\textrm{A}^\textrm{1}_\textrm{1g}$ mode of epitaxially grown, \qty{50}{\nano\meter} thick GaSe~\cite{Bissolo2025GaSe} and the $\textrm{A}_\textrm{1g}$ mode of sapphire~\cite{Thapa2017_RamanSapphire, Kadleikova2001_SapphireRaman} from Raman spectra taken on four randomly chosen spots on the sample after 2, 5, and 10 weeks of storage in vacuum, respectively. For the measurement, the spots were exposed to a laser fluence of \qty{140}{\micro\watt} (\qty{0.11}{\milli \watt \per \micro \meter^2}) for \qty{15}{\minute} integration time and kept in UHV under lab illumination in between. In case of degradation, the peak stemming from the $\textrm{A}^\textrm{1}_\textrm{1g}$ mode should broaden and decrease in intensity~\cite{Smiri2024_OxidationLongTerm}, neither of which can be observed. The average FWHM of the $\textrm{A}^\textrm{1}_\textrm{1g}$ mode after two weeks is \qty{4.3}{\cm^{-1}} and after ten weeks \qty{4.6}{\cm^{-1}}, which vary within the standard error of \qty{0.3}{\cm^{-1}} and \qty{0.1}{\cm^{-1}}, respectively. The intensity ratio I$(\textrm{A}_\textrm{1g})_{\textnormal{sapphire}}$/I$(\textrm{A}^\textrm{1}_\textrm{1g})_{\textnormal{GaSe}}$  between the Raman modes of sapphire and GaSe was plotted to disregard any effect of the measurement conditions on the absolute counts, especially in the long stability tests: The averages after two and ten weeks are \num{0.19} and \num{0.18}, respectively, which vary again within the standard errors of \num{0.02} and \num{0.003}, respectively. Fig. \ref{fig:Raman_Stability}b shows the effect of continuous illumination over \qty{1}{\hour} under an excitation power of \qty{215}{\micro\watt}, corresponding to a laser fluence of \qty{0.169}{\milli \watt \per \micro \meter^2}.  Contrary to the case under ambient conditions~\cite{Murray2025_OxidationStatistically}, the comparison of the spectra manifests no significant change, with only peaks corresponding to pristine GaSe. This is highlighted in the inset, where no trend in the evolution of the intensity of the $\textrm{A}^\textrm{1}_\textrm{1g}$ mode in GaSe relative to the $\textrm{A}_\textrm{1g}$  mode in the sapphire substrate can be observed over time, except for statistical fluctuations. The evolution is normalized to the initial ratio.
These measurements indicate that the sample degradation, as evidenced by the broadening or loss in intensity of the GaSe Raman modes, can be prevented using our UHV system, even under continuous laser illumination. This observation confirms that our approach allows continuous and reproducible study of volatile and reactive 2D materials.

\section{Conclusions and outlook}
We presented an all-UHV cluster for the epitaxial growth and optical spectroscopy of 2D semiconductors, enabling the investigation of the structural and optoelectronic properties of materials in their as-grown, pristine form. The homogeneity of the multilayer PTMC and its stoichiometric ratio were tested across an entire 2" wafer. Room-temperature Raman measurements were performed with a spatial resolution of \qty{1.2}{\um} and a spectral resolution of \qty{0.05}{\nm} for modes down to \qty{15}{\per\cm}. Photoluminescence measurements in the range between \qtyrange{20}{300}{\kelvin} were conducted using a closed-cycle cryostat to probe the excitonic properties as well as electron-phonon interactions. Vibrations induced by the cryostat lead to an asymmetric spatial resolution of \qty{8.9}{\um} and \qty{18.9}{\um} in horizontal and vertical directions, respectively. A computational method using the measured PSF of the microscope is employed to enhance the spatial resolution. In the UHV environment of the cluster with a partial oxygen pressure of \qty{5E-10}{\milli \bar}, the synthesized materials remained in a pristine state for more than 10 weeks in the vacuum environment, and for more than 1 hour under constant illumination. This allows for detailed studies aiming at understanding both the optical and structural properties without material degradation.
As a natural extension of the implemented methodology, additional characterization modes such as photoluminescence excitation spectroscopy and second-harmonic generation are currently being integrated into the platform. We further envision employing the cluster tool for the scalable growth of next-generation 2D materials combined with vacuum-transfer micro-PL and Raman spectroscopy at defined deposition stages, enabling process-correlated optical characterization without breaking vacuum.

\begin{acknowledgments}
The authors would like to thank Malte Kremser, Friedrich Sbresny, William Rauhaus, and Nathan Wilson for help with the spectroscopy setup. \\
The work was partly funded by the German Research Foundation (DFG) via Grants FI 947/7-1, FI 947/7-2, FI 947/8-1 and KO4005-9/1. In addition, we gratefully acknowledge the DFG clusters of excellence e-conversion (EXC 2089, grant number 390776260) and MCQST (EXC 2111, grant number 390814868). J.J.F acknowledges Munich Quantum Valley, which is supported by the Bavarian state government with funds from the Hightech Agenda Bavaria. J.S. gratefully acknowledges TUM-IGSSE and the Institute for Advanced Study (IAS) from Technische Universität München (TUM) for financing the focus group on “Sustainable photocatalysis using plasmons and 2D materials (SusPhuP2M)” as part of the Hans Fisher Senior Fellowships program.

\end{acknowledgments}

\section*{Conflict of Interest}

The authors declare no conflict of interest

\section*{Author Contributions}

 E.Z., G.K., and J.J.F. conceived the research. E.Z. managed it. H.R., M.D., M.B., E.Z., J.S., and A.U. built and maintained the all-UHV MBE-analytical system. The UHV microscope was designed by M.D., with input from A.T., and assembled by M.D., H.R., and J.S. J.S. developed the data acquisition software with support from J.U. M.B. carried out the MBE growth. M.D. and J.S. performed the PL and Raman measurements. P.A. fabricated and pre-characterized the quantum dot sample. M.D. and J.S. analyzed the data and discussed the results with all authors. G.K. and J.J.F., with support from E.Z., secured third-party funding. J.J.F. provided the experimental infrastructure. M.D., J.S., and E.Z. wrote the manuscript with input from all authors.

Marco Dembecki: Data curation (equal); Formal analysis (equal); Investigation (equal); Methodology (lead); Project administration (equal); Software (supporting); Validation (equal); Visualization (equal); Writing – original draft (equal); Writing – review\& editing (equal).
Jan Schabesberger: Data curation (equal); Formal analysis (equal); Investigation (equal); Methodology (supporting); Project administration (equal); Software (lead); Validation (equal); Visualization (equal); Writing – original draft (supporting); Writing – review \& editing (equal).
Michele Bissolo: Formal analysis (supporting); Investigation (equal); Methodology (supporting).
Andreas Thurn: Methodology (supporting).
Abhilash Ulhe: Methodology (supporting). 
Pavel Avdienko: Methodology (supporting).
Julius Ulrichs: Software (supporting). 
Hubert Riedl: Methodology (supporting); Project administration (supporting).
Gregor Koblmüller: Conceptualization (equal); Funding acquisition (equal); Resources(supporting); Writing – review \& editing (equal).
Eugenio Zallo: Conceptualization (equal); Funding acquisition (supporting); Methodology (supporting); Project administration (equal); Supervision (lead); Visualization (supporting); Writing – original draft (equal); Writing – review \& editing(equal).
Jonathan J. Finley: Conceptualization (equal); Funding acquisition (equal); Methodology (supporting); Project administration (equal); Resources (lead); Supervision (supporting); Writing – review \& editing (equal).

\section*{Data Availability Statement}
The data that support the findings of this study are available from the corresponding author upon reasonable request.

\bibliography{mainbib}%

\end{document}